\documentclass[graybox]{svmult}

\usepackage{helvet}         % selects Helvetica as sans-serif font
\usepackage{courier}        % selects Courier as typewriter font
\usepackage{type1cm}        % activate if the above 3 fonts are
\usepackage{makeidx}         % allows index generation
\usepackage{graphicx}        % standard LaTeX graphics tool
\usepackage{multicol}        % used for the two-column index
\usepackage[bottom]{footmisc}% places footnotes at page bottom
\usepackage{amssymb}
\usepackage{url}
\usepackage{amsmath}

\usepackage{graphicx}
\usepackage{subfigure}
\makeindex             % used for the subject index
\begin{document}

\title*{Optimal Impulse Control of SIR Epidemics over Scale-Free Networks}
\author{Vladislav~Taynitskiy$^{1}$,\ Elena~Gubar$^{1}$,\ Quanyan~Zhu$^{2}$}
\institute{Vladislav~Taynitskiy \at  St. Petersburg State University,\\  Faculty of Applied
Mathematics and Control Processes,\\ Universitetskii prospekt 35, Petergof, Saint-Petersburg, Russia, 198504; \email{tainitsky@gmail.com}
\and Elena~Gubar \at St. Petersburg State University,  St. Petersburg State University,\\  Faculty of Applied
Mathematics and Control Processes,\\ Universitetskii prospekt 35, Petergof, Saint-Petersburg, Russia, 198504; \email{e.gubar@spbu.ru}
\and Quanyan~Zhu \at Department of Electrical and Computer Engineering, Tandon School of Engineering,\\ New York University, Brooklyn, NY, USA, 11201. \email{quanyan.zhu@nyu.edu}
}
\maketitle

\abstract{
Recent wide spreading of Ransomware has created new challenges for cybersecurity over large-scale networks. The densely connected networks can exacerbate the spreading and makes the containment and control of the malware more challenging. In this work, we propose an impulse optimal control framework for epidemics over networks.  The hybrid nature of discrete-time control policy of continuous-time epidemic dynamics together with the network structure poses a challenging optimal control problem. We
leverage the Pontryagin's minimum principle for impulsive systems to obtain an optimal structure of the controller and use numerical experiments to corroborate our results.
}

\section{Introduction}\label{intro}

Malware spreading becomes a more prevalent issue recently as the number of devices and their connections grow exponentially. Many devices that are connected to the Internet do not have strong protections, and they contain cyber vulnerabilities that create a fast spreading of malware over large networks. A higher level of connectivity of the network is often desired for information spreading. However, in the context of malware, the high connectivity can exacerbate the spreading and makes the containment and control of the malware more challenging. One example is the recent Ransomware \cite{Ransomware,Luo} that spreads over the Internet with the objective to   
lock the files of a victim using cryptographic techniques and demand a ransom payment to decrypt them. The worldwide spread of  WannaCry ransomware has affected more than 200,000 computers across 150 countries and caused billions of dollars of damages. 
Hence it is critical to take into account the network structure when developing control policies to control the infection dynamics. 

In this paper, we investigate a continuous-time Susceptible-Infected-Recovered
(SIR) epidemic model over large-scale networks. The malware
control mechanism is to patch an optimal fraction of the
infected nodes at discrete points in time. Such mechanism is
also known as an optimal impulse controller. The hybrid nature of discrete-time control policy of continuous-time epidemic
dynamics together with the network structure poses a challenging optimal control problem. We
leverage the Pontryagin's minimum principle for impulsive
systems to obtain an optimal structure of the controller and
use numerical experiments to demonstrate the computation of
the optimal control and the controlled dynamics. This work
extends the investigation of previous related works \cite{Zhu,Gubar,Taynitskiy} to
a new paradigm of coupled epidemic models and the regime
of optimal impulsive control. % This security concern requires 

%Recent advances in information and communication technologies
%(ICT) have enabled the interconnections of an increasing
%number of devices into the Internet [9]. One critical
%emerging technology is the Internet of Things (IoTs) that
%connects physical objects and devices with communication
%networks to create a next-generation Internet that can interact
%with smart things [12]. However, the cyber vulnerabilities
%of the IoT allow the spreading of malware which creates
%a significant security concern. One example is the recent
%malware Mirai which spreads through infected surveillance
%cameras and launches a distributed denial-of-service (DDoS)
%attack that has disrupted the Internet services of US in 2016.
%Heterogeneity is one prominent feature of IoTs as a number
%of diverse devices can communicate through different networks.
%Therefore, the emerging IoT networks can be infected
%by heterogeneous malware that exploits different types of
%system vulnerabilities. Hence it is of paramount importance
%to understand the infection dynamics and develop mechanisms
%to contain the spreading through periodic monitoring, patching
%and diagnostics in the emerging network.
%In this paper, we investigate a continuous-time SusceptibleInfected-Recovered
%(SIR) epidemic model for heterogeneous
%populations over a large network of devices. 

The rest of the paper is organized as follows. Section II
presents the controlled SIR mathematical model. Section III
shows the structure of optimal control policies. Section IV
presents numerical examples. Section V concludes the paper.

\section{The model}\label{model}
In this section, we formulate  a model of spreading of malware in the network of $N$ nodes use the  modification of classical SIR model. As in previous works \cite{Gubar}, \cite{Taynitskiy:2016} two different forms of malware with different strengths spread over the network simultaneously, we denote them as $M_1$ and $M_2$.  We also assume that a structure of population  is described by the scale free network  \cite{Vespignani}, \cite{Porokhnyavaya}.  Normally, as SIR model points, all nodes in the population are  divided into three groups: \textit{Susceptible} $(S)$, \textit{Infected }$(I)$ and \textit{Recovered} $(R)$, \cite{Altman}.  \textit{Susceptible} is a group of nodes which are not infected by any malware, but may be invaded by any forms of virus. The \textit{Infected} nodes are those that have been attacked by the virus and the \textit{Recovered} is a group of recovered nodes. In modified model subgroup of Infected nodes also is brunched into two subgroups $I_1$ and $I_2$, where nodes in $I_i$ are infected by malware $M_i, i=1,2$ respectively. We formulate the epidemic process as a system of nonlinear differential equations, where  $n_S$, $n_{M_1}$, $n_{M_2}$ and $n_R$ correspond to the number of susceptible, infected and recovered nodes, respectively. In current model the connections between nodes are described by the scale-free network, then we will use the following notation: $S_k(t)$ and $R_k(t)$ are fractions of \textit{Susceptible} and \textit{Recovered} nodes with degree $k$ at time moment $t$, $I^1_k(t), I^2_k(t)$ are fractions of \textit{Infected} nodes with degree $k$. At each time moment $t\in [0,T]$  the number of nodes is constant and equal $N$, and the following condition $S_k(t)+ I^1_k(t)+ I^2_k(t)+ R_k(t)=1$ is satisfied. The process of spreading is defined by the system of ordinary differential equations:

\begin{equation}\label{SIR_main}\begin{array}{l}
\displaystyle\frac{dS_k}{dt}=-\delta_{1k} S_kI^1_k\Theta_1-\delta_{2k} S_kI^2_k\Theta_2;\\
\displaystyle\frac{dI^1_k}{dt}=\delta_{1k} S_k I^1_k\Theta_1 -\sigma^1_k I^1_k;\\
\displaystyle\frac{dI^2_k}{dt}=\delta_{2k} S_k I^2_k\Theta_2 -\sigma^2_k I^2_k,\\
\displaystyle\frac{dR_k}{dt}=\sigma^1_k I^1_k+\sigma^2_k I^2_k,
\end{array}\end{equation}
 where $\delta_{ik}(k)$ is the infections rate for the first type of malware $i$ if a susceptible node has a contact with infected node with the degree $k$, $\sigma^i_k$ is recovery rate.

We consider the graph generated by using the algorithm devised in \cite{barabasi}. We start from a small number $m_0$ of disconnected nodes; every time step a new node is added, with $m$ links that are connected to an old node $i$ with $k_i$ links according to the probability $k_i/\sum_j k_j$. After iterating this scheme a sufficient number of times, we obtain a network composed by $N$ nodes with connectivity distribution $P(k) \approx k^{-3}$ and average connectivity $\langle k\rangle=2m$. In this work we take $m=4$.

At the initial time moment $t=0$, the most number of nodes belong to Susceptible group and only a small fraction of Infected by malwares $M_1$ or $M_2$. Initial state for system (\ref{SIR_main}) is $0<S_k(0)< 1$, $0<I^1_k(0)< 1$, $0<I^2_k(0)< 1$, $R(0)=1-S_k(0)-I^1_k(0)-I^2_k(0)$. Analogously with  \cite{Fu}, \cite{Vespignani} we define parameter $\Theta_i(t)$ as

\begin{equation}\label{theta_general}
\Theta_i(\lambda_i)=\sum\limits_{k'}\frac{\delta_{ik} P(k'|k)I^i_{k'}}{k'}, \ i=1,2,
\end{equation}
where $\delta_{ik}$ denotes the infectivity of a node with degree $k$ and $\lambda_i=\delta_{ik}/\sigma^i_k$ an effective spreading rate.  $P(k'|k)$ describes  the probability of a node with degree
$k$ pointing to a node with degree $k'$, and $P(k'|k)=\frac{k' P(k')}{k'}$, where $\langle k\rangle=\sum\limits_{k'}k P(k)$. For scale-free node distribution $P(k)=C^{-1}k^{-2-\gamma},\ 0<\gamma \le 1$, where $C=\zeta (2+\gamma)$ is Riemann's zeta function, which provides an appropriate normalization constant for sufficiently large networks. In the SF model considered here, we have a connectivity distribution $P(k)=2m^2/k^{-3}$, where $k$ is approximated as a continuous variable. According to \cite{Vespignani} we can rewrite (\ref{theta_general}) as

\begin{equation}\label{theta}
\Theta_i(\lambda_i)=\frac{e^{-1/m\lambda_i}}{m\lambda_i}, \ i=1,2.
\end{equation}

\section{Impulse control problem}\label{impulse_control}

Previously it was shown in \cite{Vespignani} a small fraction of the infected nodes might be survived on small segments of the network and can provoke new waves of epidemics. This cycled process recalls
 the behavior the virus of influenza which causes a seasonally periodic epidemic, \cite{Agur}.  Hence the control of the epidemic process can be formulated as an impulse control problem in which a series of impulses of antivirus patches are designed to reduce the periodically incipient zones of infected nodes.  We extend the model (\ref{SIR_main})  to present an impulse control problem for episodic attacks of the malware and obtain the optimal strategy of application of antivirus software to damp the spreading of malware at discrete time moments.

 We suppose that  impulses occur at  time  $\tau^i_{k,1},\ldots,\tau^i_{k,q_i(k)}$, where $q_i(k)$ describes the number of launching of impulse controls for nodes with  $k$ degrees,  index $i$ indicates the type of malware.  We also assume that  on the time intervals $(\tau^i_{k,j}, \tau^i_{k,j+1}]$ system (\ref{SIR_main}) describes the behavior of malware  in the network. We have reformulated epidemic model  to describe the situation with two types of malware  for all time periods except the sequence of times $\tau_{k,j}^{i+}$, $j=1,\ldots, q_i(k)$, $i=1,2$. Additionally, we set $S(\tau_{k,j}^{i})=S(\tau_{k,j}^{i-})$, $I_1(\tau_{k,j}^{i})=I_1(\tau_{k,j}^{i-})$,  $I_2(\tau_{k,j}^{i})=I_2(\tau_{k,j}^{i-})$, $R(\tau_{k,j}^{i})=R(\tau_{k,j}^{i-})$.

The system after activation of impulses at time moment $\tau_{k,j}^{i+}$ is:
\begin{equation}\label{SIR_impulse}\begin{array}{l}
S_k(\tau_{k,j}^{i+})=S_k(\tau_{k,j}^{i}),\\
{I^1_k}({\tau_{k,j}^{i+}})=I^1_k(\tau_{k,j}^{i}) -\nu^1_k(\tau_{k,j}^{i}),\\
{I^2_k}({\tau_{k,j}^{i+}})=I^2_k(\tau_{k,j}^{i}) -\nu^2_k(\tau_{k,j}^{i}),\\
{R_k}({\tau_{k,j}^{i+}})={R_k}(\tau_{k,j}^{i}) +\nu^1_k(\tau_{k,j}^{i}) +\nu^2_k(\tau_{k,j}^{i}).
\end{array}\end{equation}

Variables $\nu^i_k=({\nu^i_{k,1}},\ldots,{\nu^i_{k,q_i(k)}})$, $i=1,2,$ correspond to control impulses applied at the discrete time moments $\tau_{k,1},\ldots,\tau_{k,q_i(k)}$ and represent the fraction of recovered nodes. Let be $\nu^i_{k,j}={c^i_{k,j}}\delta (t-{\tau^i_{k,j}})$, where $\delta (t-{\tau^i_{k,j}})$ is   Dirac function, ${c^i_{k,j}}\in[0, {\overline{u}^i_{k,j}}]$ is the value of impulse, leads to changes of the dynamical system, ${\overline{u}^i_{k,j}}$ is the maximum value for control \cite{Agur}.

\textbf{Functional:} the objective function of the combined system (\ref{SIR_impulse}) is represented by the aggregated costs on the time interval $[0,T]$ including the costs of control impulses.  The aggregated costs for continuous system (\ref{SIR_main}) are defined as follows:  at time moment  $t\neq {\tau^i_{k,j}}, \ j=1,\ldots,q_i(k)$, $i=1,2$, we have the costs from infected nodes $f^1_k(I^1_k(t))$ and $f^2_k(I^2_k(t))$.  Functions $f^i_k(\cdot)$ are non-decreasing and twice-differentiable, such that $f^i_k(0)=0$, $f^i_k(I^i_k(t))>0$ for $I^i_k(t)>0$ with $t\in ({\tau^i_{k,j-1}}, {\tau^i_{k,j}}]$.  For system (\ref{SIR_impulse}), we define the treatment costs as  functions $h^i_k(\nu^i_{k,j}({\tau^{i+}_{k,j}}))$, $j=1,\ldots,{q_i(k)}$, where $h^i_k(\nu^i_{k,j}({\tau^{i+}_{k,j}}))>0$, $\nu^i_{k,j}({\tau^{i+}_{k,j}})>0$ for $i=1,2$. Functions $g(R_k(t))$ are non-decreasing and capture the benefit rates from Recovered nodes. The aggregated system costs are defined by the  functional:

\begin{equation}\label{functional_J}\begin{array}{l}
J=\sum\limits_{k\in \mathbb{N}} [ \int_0^T
f^1_k(I^1_k(t))+f^2_k(I^2_k(t))-g(R(t))dt+\\
\hskip 40pt \sum\limits_{j=1}^{q_1(k)} h^1_k(\nu^1_{k,j}({\tau^1_{k,j}}))+\sum\limits_{j=1}^{q_2(k)} h^2_k(\nu^2_{k,j}({\tau^2_{k,j}}))].
\end{array}
\end{equation}

\section{The structure of impulse control}\label{impulse_control}

According to principle maximum in impulse form \cite{Blaquiere}, \cite{Chahim}, \cite{Dykhta}, \cite{Taynitskiy}  we write Hamiltonian for dynamic system (\ref{SIR_main})
\begin{equation} \begin{array}{l}
H^0_k(t)=-(f^1_k(I^1_k(t))+f^2_k(I^2_k(t))-g(R_k(t)))+(\lambda_{I^1_k}(t)-\lambda_{S_k}(t))\delta_{1k}S_k(t) I^1_k(t)\Theta_1(t)+\\
\hskip 35pt (\lambda_{I^2_k}(t)-\lambda_{S_k}(t))\delta_{2k} S_k(t) I^2_k(t)\Theta_2(t)+(\lambda_{R_k}-\lambda_{I^1_k})\sigma^1_k I^1_k+(\lambda_{R_k}-\lambda_{I^2_k})\sigma^2_k I^2_k;
\end{array}\end{equation}

and construct adjoint system as follows:

\begin{equation} \begin{array}{l}
\displaystyle\dot{\lambda}_{S_k}(t)=(\lambda_{S_k}(t)-\lambda_{I^1_k}(t))\delta_{1k}I^1_k(t)\Theta_1(t)+(\lambda_{S_k}(t)-\lambda_{I^2_k}(t))\delta_{2k}I^2_k(t)\Theta_2(t);\\
\displaystyle\dot{\lambda}_{I^1_k}(t)=\frac{df^1_k(I^1_k(t))}{dI^1_k}+(\lambda_{S_k}(t)-\lambda_{I^1_k}(t))\left(\delta_{1k}S_k(t)\Theta_1(t)+\frac{(\delta_{1k})^2S_k(t) I^1_k(t) P(k)}{\langle k\rangle}\right)+\\
\hskip 35pt (\lambda_{I^1_k}-\lambda_{R_k})\sigma^1_k;\\
\displaystyle\dot{\lambda}_{I^2_k}(t)=\frac{df^2_k(I^2_k(t))}{dI^2_k}+(\lambda_{S_k}(t)-\lambda_{I^2_k}(t))\left(\delta_{2k}S_k(t)\Theta_2(t)+\frac{(\delta_{2k})^2S_k(t) I^2_k(t) P(k)}{\langle k\rangle}\right)+\\
\hskip 35pt (\lambda_{I^2_k}-\lambda_{R_k})\sigma^2_k;\\
\displaystyle\dot{\lambda}_{R_k}(t)=-\frac{dg(R_k(t)}{dR_k},
\end{array}\end{equation}
with transversality conditions $\lambda_{S_k}(T)=\lambda_{I^1_k}(T)=\lambda_{I^2_k}(T)=\lambda_{R_k}(T)=0$.

Following the  maximum principle for impulse control (see \cite{Blaquiere}, \cite{Sethi}, \cite{Chahim}), we formulate necessary optimality conditions as in Theorem \ref{thm_1}.

\begin{theorem}\label{thm_1}
Let $(x^*,N,\tau_i^{j*},\nu_i^*)$, $i=1,2,$ be an optimal solution for the impulse control problem. Then, there exists an adjoint vector function $\lambda(t)=(\lambda_S(t),\lambda_{I_{1}}(t),\lambda_{I_{2}}(t)$, $\lambda_{R}(t))$ such that the following conditions hold:
\begin{equation}\begin{array}{l}
\dot{\lambda}_x(t) =-\frac{\partial H_0}{\partial x}(x^*(t),\lambda(t),t),
\end{array}\end{equation}
where $x(t)={S(t),I_1(t),I_2(t),R(t)}$.

At the impulse or jump points, it holds that
\begin{equation}\label{thm_cond_1}\begin{array}{l}
\frac{\partial H_i^c}{\partial \nu_i}(x^*(\tau_i^{j*-}),\nu_i,\lambda(\tau_i^{j*+}),\tau_i^{j*})(\nu_i^j-\nu_i^{j*})\ge 0,\\ \\\end{array}\end{equation}
\begin{equation}\label{thm_cond_2}\begin{array}{l}
\lambda_x(\tau_i^{j*+})-\lambda_x(\tau_i^{j*-})=\frac{\partial H_i^c}{\partial x}(x^*(\tau_i^{j*-}),\nu_i^{j*},\lambda(\tau_i^{j*+}),\tau_i^{j*}),\\ \\\end{array}\end{equation}
{\begin{equation}\label{thm_cond_3}\begin{array}{l}
H_0(x^*(\tau_i^{j*+}),\lambda(\tau_i^{j*+}),\tau_i^{j*})-H_0(x^*(\tau_i^{j*-}),\lambda(\tau_i^{j*-}),\tau_i^{j*})- \bigskip\\
\hskip 1pt -\frac{\partial H_i^c}{\partial \tau_i^j}(x^*(\tau_i^{j*-}),\nu_i^{j*},\lambda(\tau_i^{j*+}),\tau_i^{j*})

\begin{cases}
>0 & \text{for $\tau_i^{j*}=0$},\\
=0 & \text{for $\tau_i^{j*}\in(0,T)$,}\\
<0 & \text{for $\tau_i^{j*}=T$.}\\
\end{cases}
\end{array}\end{equation}}

For all points in time at which there is no jump, i.e. $t\neq \tau_j\ (j=1, \ldots, k_i)$, it holds that
\begin{equation}\begin{array}{l}
\frac{\partial H_j^c}{\partial \nu_j}(x^*(t),0,\lambda(t),t)\nu_j\le 0,
\end{array}\end{equation}
with the transversality condition $\lambda(T)=0.$
\end{theorem}

Hamiltonian in impulsive form is
\begin{equation} \begin{array}{l}
H^c_k(\tau_{k,j}^{1+})=-h^1_k(\nu^1_{k,j}(\tau_{k,j}^{1+}))+(\lambda_{R_k}(\tau_{k,j}^{1+})-\lambda_{I^1_k}(\tau_{k,j}^{1+}))\nu^1_{k,j}(\tau_{k,j}^{1+});\\
H^c_k(\tau_{k,j}^{2+})=-h^2_k(\nu^2_{k,j}(\tau_{k,j}^{2+}))+(\lambda_{R_k}(\tau_{k,j}^{2+})-\lambda_{I^2_k}(\tau_{k,j}^{2+}))\nu^2_{k,j}(\tau_{k,j}^{2+}).
\end{array}\end{equation}

Here we assume that for each type of malwares $M_1$ and $M_2$  and for each $k$ we have own set of control impulses $\nu^1_k=({\nu^1_{k,1}},\ldots,{\nu^1_{k,q_1(k)}})$ and $\nu^2_k=({\nu^2_{k,1}},\ldots,{\nu^2_{k,q_2(k)}})$.

Adjoin system for system(\ref{SIR_impulse}) is ($i=1,2$):
\begin{equation} \begin{array}{l}
\displaystyle\lambda_{S_k}(\tau_{k,j}^{i+})=\lambda_{S_k}(\tau_{k,j}^{i});\\
\displaystyle\lambda_{I^1_k}(\tau_{k,j}^{i+})=\lambda_{I^1_k}(\tau_{k,j}^{i});\\
\displaystyle\lambda_{I^2_k}(\tau_{k,j}^{i+})=\lambda_{I^2_k}(\tau_{k,j}^{i});\\
\displaystyle\lambda_{R_k}(\tau_{k,j}^{i+})=\lambda_{R_k}(\tau_{k,j}^{i}).
\end{array}\end{equation}

Here is the conditions for  $\Delta_i$ for each $I^i_k$ from the theorem \ref{thm_1}:
\begin{equation} \begin{array}{l} \label{Delta_1}
\displaystyle\Delta_1=f^1_k(I^1_k(\tau_{k,j}^{1}))-f^1_k(I^1_k(\tau_{k,j}^{1+}))-g(R_k(\tau_{k,j}^{1}))+g(R_k(\tau_{k,j}^{1+}))+\\
\hskip 23pt \displaystyle c^1_{k,j}[\frac{d g(R_k(\tau_{k,j}^{1+}))}{d R_k(\tau_{k,j}^{1+})} + \frac{df^1_k(I^1_k(\tau_{k,j}^{1+}))}{dI^1_k(\tau_{k,j}^{1+})}]+ 2\sigma^1_k c^1_{k,j}(\lambda_{R_k}(\tau_{k,j}^{1+})-\lambda_{I^1_k}(\tau_{k,j}^{1+}))\\

\hskip 23pt \displaystyle\delta_{1k}S_k(\tau_{k,j}^{1})c^1_{k,j}(\lambda_{S_k}(\tau_{k,j}^{1})-\lambda_{I^1_k}(\tau_{k,j}^{1}))[2\Theta_1(\tau_{k,j}^{1+})+\frac{\delta_{1k} P(k)}{\langle k\rangle}(1+I^1_k(\tau_{k,j}^{1})-c^1_{k,j})].
\end{array}\end{equation}

Here is the conditions for  $\Delta$ for each $\Delta$ for each $I^2_k$ from theorem \ref{thm_1}:
\begin{equation} \begin{array}{l} \label{Delta_2}
\displaystyle\Delta_2=f^2_k(I^2_k(\tau_{k,j}^{2}))-f^2_k(I^2_k(\tau_{k,j}^{2+}))-g(R_k(\tau_{k,j}^{2}))+g(R_k(\tau_{k,j}^{2+}))+\\
\hskip 23pt \displaystyle c^2_{k,j}[\frac{d g(R_k(\tau_{k,j}^{2+}))}{d R_k(\tau_{k,j}^{2+})} + \frac{df^2_k(I^2_k(\tau_{k,j}^{2+}))}{dI^2_k(\tau_{k,j}^{2+})}]+ 2\sigma^2_kc^2_{k,j}(\lambda_{R_k}(\tau_{k,j}^{2+})-\lambda_{I^2_k}(\tau_{k,j}^{2+}));\\
\hskip 23pt \displaystyle\delta_{2k}S_k(\tau_{k,j}^{2})c^2_{k,j}(\lambda_{S_k}(\tau_{k,j}^{2})-\lambda_{I^2_k}(\tau_{k,j}^{2}))[2\Theta_2(\tau_{k,j}^{2+})+\frac{\delta_{2k} P(k)}{\langle k\rangle}(1+I^2_k(\tau_{k,j}^{2})-c^2_{k,j})].
\end{array}\end{equation}

According to Theorem \ref{thm_1} at time $\tau^{i}_{k,j} \in (0,T)$ $\Delta_i$ should be equal to zero.  Therefore, we deal with  two different problems: firstly, if the intensity of impulses $c^{i}_{k,j}$ are  fixed, then from (\ref{Delta_1}) and (\ref{Delta_2}),  we can find the optimal time  $\tau^{i*}_{k,j}$ of using impulses; secondly, if the sequence of time $\tau^{i}_{k,j}$  are fixed, then we obtain  the optimal level of the intensity of impulses $c^{i*}_{k,j}$, $j=1,\ldots,q_i$, $i=1,2$.

\section{Numerical simulations}\label{numerical_simulations}

In this paragraph we present numerical experiments to depict theoretical results and to study the   behavior of malwares  and show the application of control impulses. Here  we use the following set of the initial states and values of parameters of the system (\ref{SIR_main}): initial system states and parameters are $S_k(0)=0.4,$ $I^1_k(0)=0.3,$ $I^2_k(0)=0.2$ and $R_k(0)=0$, spreading rates are $\delta_{1k}=0.075k$, $\delta_{2k}=0.1k$, self-recovery rates are $\sigma^1_k=0.0005k$ and $\sigma^1_k=0.0003k$. We set costs functions for infectious subgroups as $f^i_k(I^i_k(t))=A^i_k  I^i_k(t)$ with coefficients $A^1_k=2k$, $A^2_k=3k$ and treatment costs functions as $h^i_k(\nu^i_{k,j}(\tau^{i+}_{k,j})=B^i_k   c^i_{k,j} I^i_k(\tau^{i+}_{k,j})$, where coefficients are equal to  $B^1_k=3k$, $B^2_k=4k$, $c^1_{k,j}=0.1$, $c^2_{k,j}=0.08$ for $i=1,2$, utility function is $g(R_k(t))=0.1 R_k(t)$.

\begin{center}
\begin{figure}[h!]
%\begin{minipage}{0.68\linewidth}
\centering \includegraphics[width=0.68\linewidth]{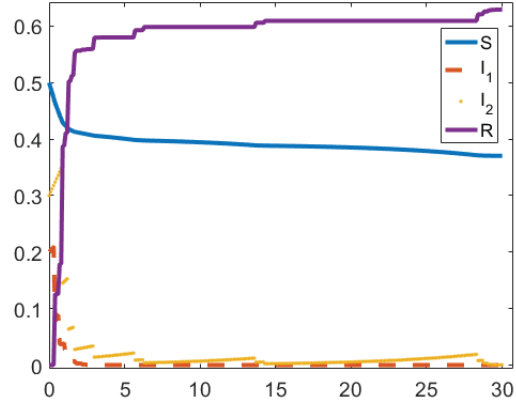}
\caption{Evolution of the system in Case 1. Number of links: $k=4$, spreading rates: $\delta_{1k}=0.075k$ and $\delta_{2k}=0.1k$. }
\label{SIR_37_65.eps}
\end{figure}
\end{center}

\begin{center}
\begin{figure}[h!]
%\begin{minipage}{0.68\linewidth}
\centering\includegraphics[width=0.68\linewidth]{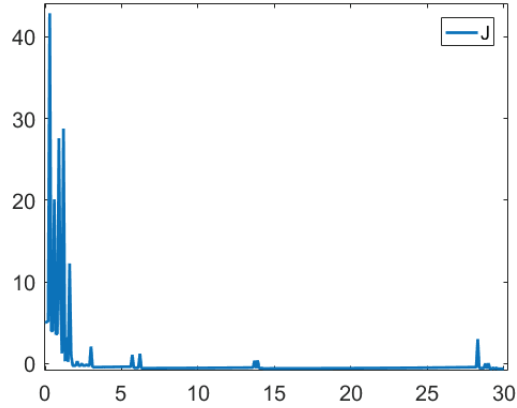}
\caption{Aggregated system costs are equal to $J=37.65$.
\vspace{\baselineskip}}
\label{J_37_65.eps}
\end{figure}
\end{center}

\textbf{Case 1.} In case 1 we present the initial example of the behavior of the system and aggregated system costs if an average number of links between $i$-th node and its neighbors is $k=4$. Figs. \ref{SIR_37_65.eps}-\ref{J_37_65.eps} show the spreading on two modification of malwares and corresponding total system costs.

%\begin{figure}[ht!]
%\vspace{1ex} 
%\centering 
%\subfigure[]{
%\includegraphics[width=1.1\linewidth]{SIR_37_65.eps}\caption{Evolution of the system in Case 1. Number of links: $k=4$, spreading rates: $\delta_{1k}=0.075k$ and $\delta_{2k}=0.1k$. }
%\label{SIR_37_65.eps}}
%\subfigure[]{
%\includegraphics[width=1.1\linewidth]{J_37_65.eps}
%\caption{Aggregated system costs are equal to $J=37.65$.}
%\label{J_37_65.eps}}
%\end{figure}

%\end{minipage}
%\hfill

%
%\begin{minipage}{0.68\linewidth}
%\includegraphics[width=1.1\linewidth]{J_37_65.eps}
%\caption{Aggregated system costs are equal to $J=37.65$.
%\vspace{\baselineskip}}
%\label{J_37_65.eps}
%\end{minipage}
%\end{figure}
%\end{center}

%\begin{figure}[ht!]
%\vspace{1ex} \centering \subfigure[]{
%\includegraphics[width=0.48\linewidth]{SIR_37_65.eps} \label{SIR_37_65.eps}
%%\hspace{3ex}
%\subfigure[]{
%\includegraphics[width=0.48\linewidth]{J_37_65.eps} \label{J_37_65.eps}}
%\caption{\subref{SIR_37_65.png} Evolution of the system in Case 1. Number of links: $k=4$, spreading rates: $\delta_{1k}=0.075k$ and $\delta_{2k}=0.1k$. \subref{J_37_65.eps} Aggregated system costs are equal to $J=37.65$.}
%\end{figure}

 Aggregated system costs in this case are equal to $J=37.65$. By applying the control impulses at discrete time moments we received that an amount of impulses are equal to $p_1(4)=37$ and $p_2(4)=49$.

\textbf{Case 2.} In this experiment we use the same parameters for initial data, but in contrast tot xase 1 an average number on neighbors is equal $k=7$.
 In this case, we obtain that the aggregated costs are $J=73.93$, and an amount of impulses are equal to $p_1(7)=29$ and $p_2(7)=44$. We may notice that increasing the number of neighbor links increases the costs of the system. Since there are less nodes with connectivity $k=7$ which is more than average connectivity $\langle k\rangle=4$, we need less impulse treatment to vaccinate the network, thereby if we apply control to more connected nodes we reduce the costs of treatment.

\begin{figure}[ht!]
\vspace{1ex} \centering \subfigure[]{
\includegraphics[width=0.68\linewidth]{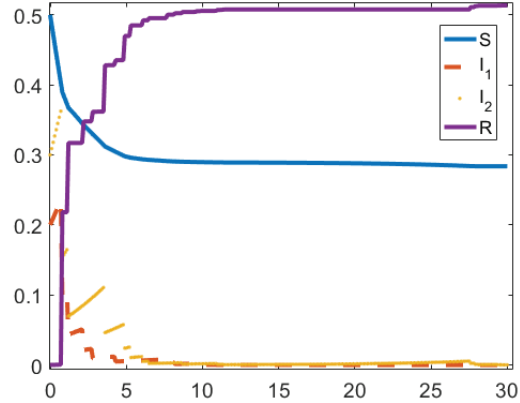} \label{SIR_73_93.eps}
%\hspace{3ex
}
\subfigure[]{
\includegraphics[width=0.68\linewidth]{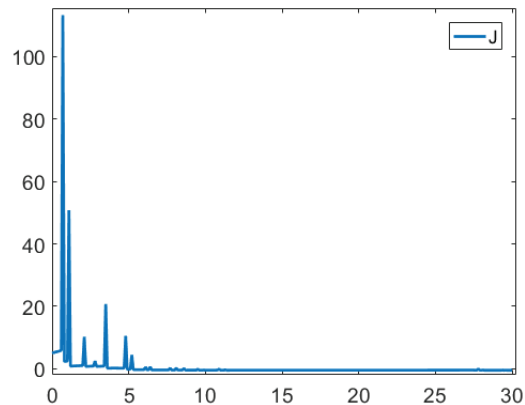} \label{J_73_93.eps}}
\caption{\subref{SIR_73_93.eps} Evolution of the system in Case 2. Number of links: $k=7$, spreading rates: $\delta_{1k}=0.075k$ and $\delta_{2k}=0.1k$. \subref{J_73_93.eps} Aggregated system costs are equal to $J=73.93$.}
\end{figure}

\textbf{Case 3.} In case 3, by using the same initial set of data we variate the spreading rate for malwares and consider  $\delta_{1k}=0.075k$ and $\delta_{2k}=0.1k$. Here we receive that  the aggregated costs are $J=122.27$ and a number of impulses are $p_1(4)=43$ and $p_2(4)=55$, then increasing the spreading rates are leading to increasing aggregated costs and number of impulses which are needed to heal the network.

\begin{figure}[ht!]
\vspace{1ex} \centering \subfigure[]{
\includegraphics[width=0.68\linewidth]{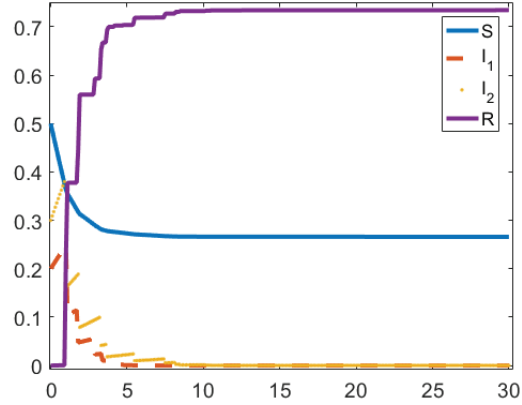} \label{SIR_122_27.eps}
%\hspace{3ex
}
\subfigure[]{
\includegraphics[width=0.68\linewidth]{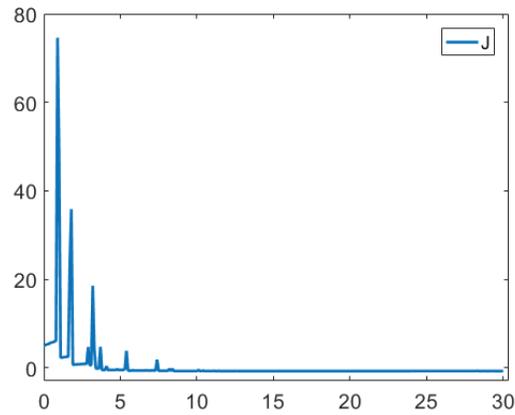} \label{J_122_27.eps}}
\caption{\subref{SIR_122_27.eps} Evolution of the system in Case 3. Number of links: $k=4$, spreading rates: $\delta_{1k}=0.1k$ and $\delta_{2k}=0.2k$. \subref{J_122_27.eps} Aggregated system costs are equal to $J=122.27$.}
\end{figure}

\section{conclusion}
This work addresses the spreading of cyber threats over large-scale networks by investigating the optimal control policies in the impulsive form for SIR-type of epidemics over scale-free networks. We have applied the impulse optimal control framework to the epidemics over networks to devise impulsive protection policies to mitigate the impact of the epidemics and contain the spreading of the malware. With the application of the maximum principle, we have obtained the structure of the optimal control impulses where actions are taken at discrete-time moments. We have corroborated the obtained results using numerical examples.

%This paper considers the case of application of optimal control in the impulsive form to protect the population against the epidemics of heterogeneous viruses. We have established the optimal control framework for epidemic models of two coexisting types of virus for heterogeneous populations, where control impulses are actioned at discrete time moments. We have obtained the structure of the optimal impulse controller and used the numerical examples to corroborate the results.

\begin{acknowledgement}
The work of the second author was supported by the research grant ``Optimal Behavior in Conflict-Controlled Systems'' (17-11-01079) from Russian Science Foundation.
\end{acknowledgement}


\begin{thebibliography}{99}

\bibitem{Ransomware} Kharraz, A., Robertson, W., Balzarotti, D., Bilge, L., Kirda, E, \newblock Cutting the gordian knot: A look under the hood of ransomware attacks. \newblock In \emph{International Conference on Detection of Intrusions and Malware, and Vulnerability Assessment}, pp. 3-24. Springer,  2015.

\bibitem{Blaquiere} Blaqui$\grave{e}$re A. \newblock Impulsive Optimal Control
with Finite or Infinite Time Horizon. \newblock \emph{Journal Of Optimization Theory And Applications.}, vol. 46.  1985.

\bibitem{Chahim} Chahim, M., Harti, R., Kort, P. \newblock A tutorial on the deterministic Impulse Control Maximum Principle: Necessary and sufficient optimality conditions.\newblock \emph{European Journal of Operational Research.}, 219, 18--26, 2012.

\bibitem{Dykhta} Dykhta V. A., Samsonyuk O. N. \newblock A maximum principle for smooth optimal impulsive control problems with multipoint state constraints. \newblock \emph{ Computational Mathematics and Mathematical Physics.},
 49,  942--957,  2009.

\bibitem{Taynitskiy} Taynitskiy V., Gubar E., Zhu Q. Optimal Impulse Control of Bi-Virus SIR Epidemics with Application to Heterogeneous Internet of Things. \emph{Constructive Nonsmooth Analysis and Related Topics. Abstracts of the International Conference. Dedicated to the Memory of Professor V.F. Demyanov}. p. 113-116, 2017.



%\bibitem{Gubar}
%\ab{Gubar E., Zhu Q.} Optimal control of influenza epidemic model with virus mutations. In:  Proc. European Control Conference (ECC), IEEE. p. 3125--3130, 2013.

\bibitem{barabasi}
%Barab{\'a}si A.-L., Albert R. Science 286, 509, 1999.
Barabási, A. L., Albert, R. Emergence of scaling in random networks. science, 286(5439), pp. 509-512, 1999.

\bibitem{Zhu} Gubar E., Zhu Q., Taynitskiy V. Optimal Control of Multi-strain Epidemic Processes in Complex Networks. \emph{Game Theory for Networks. GameNets 2017. Lecture Notes of the Institute for Computer Sciences, Social Informatics and Telecommunications Engineering, vol 212}. Springer, Cham. p. 108--117, 2017.

\bibitem{Porokhnyavaya}
Gubar E., Kumacheva S., Zhitkova E., Porokhnyavaya O. Impact of Propagation Information in the Model of Tax Audit. \emph{Recent Advances in Game Theory and Applications. "Static and Dynamic Game Theory: Foundations and Applications"}.  Switzerland. p. 91--110, 2015.

\bibitem{Fu}Fu X.,   Small M.,   Walker D. M.,  Zhang H. Epidemic dynamics on scale-free networks with piecewise linear infectivity and immunization. \emph{Phys. Rev. E. 77, 3, 036113}, 2008.

\bibitem{Vespignani}Pastor-Satorras R.,  Vespignani A. Epidemic spreading in scale-free networks. \emph{Phys. Rev. Lett. 86, 14, 3200,} 2001.

\bibitem{Agur}Agur Z., Cojocaru L., Mazor G., Anderson R. M., Danon Y. L. Pulse mass measles vaccination across age cohorts. \emph{Proceedings of the National Academy of Sciences of the United States of America.} 90:\penalty 0 11698--11702, 1993.

\bibitem{Luo} Luo X, Liao Q. Ransomware: A new cyber hijacking threat to enterprises. \emph{Handbook of research on information security and assurance.} 2009:1-6.

\bibitem{masuda2006multi}
Masuda N., Konno N. Multi-state epidemic processes on complex networks. \emph{J. Theor. Biol.243, 1, 64--75}, 2006.

\bibitem{Sethi} Sethi, S.P., Thompson, G.L. \newblock Optimal Control Theory: Applications to Management Science and Economics. \newblock Springer, Berlin, 2006.
  \bibitem{Taynitskiy:2016} Taynitskiy V. A., Gubar E. A., Zhitkova E. M.\newblock  Optimization of protection of computer networks against malicious software, \emph{Proc. of International Conference "Stability and Oscillations of Nonlinear Control Systems" (Pyatnitskiy's Conference)}, 2016.
\bibitem {Zaccour} Zaccour, G., Reddy, P., Wrzaczek, S.\newblock Quality effects in different advertising models - An impulse control approach. \emph{ European Journal of Operational Research.}, 255, 984--995, 2016.

    \bibitem{Leander} Leander R.,   Lenhart S. and Vladimir Protopopescu V. \newblock Optimal control of continuous systems with impulse controls. \newblock \emph{Optimal Control Applications and Methods Optim. Control Appl. Meth.}, 36:\penalty 535 --549, 2015.

\bibitem{Gubar} Gubar, E., Zhu, Q.\newblock Optimal Control of Influenza Epidemic Model with Virus Mutations.\newblock  \emph{ Proceedings 12th Biannual European Control Conference.} IEEE Control Systems Society. 3125--3130, 2013.
\end{thebibliography}
\end{document}